\documentclass[aps,prb,reprint,groupedaddress, showpacs]{revtex4-1}
\usepackage{txfonts}
\usepackage{braket}
\usepackage{graphicx}
\usepackage{bm}
\usepackage[]{hyperref}
\begin{document}
\title{Spontaneous loop-spin current with topological characters in the Hubbard model}
\author{Shimpei Goto, Keisuke Masuda, and Susumu Kurihara}
\affiliation{Department of Physics, Waseda University, Shinjuku, Tokyo 169-8555, Japan}
\date{\today}
\begin{abstract}
We find a state characterized by a spontaneous loop-spin current and a single-particle gap in the Hubbard model within the variational cluster approach.
This state exists for arbitrarily small interaction in a half-filled honeycomb lattice.
Moreover, from the calculations of the topological invariants for the interacting system, it is shown that this gapped state has nontrivial topological
characters; this state is the topological Mott insulating state. This result implies the ubiquity of topological Mott insulating phases.
\end{abstract}
\maketitle
\section{Introduction}
The loop-current phase\cite{PhysRevB.37.3774, PhysRevB.55.14554, PhysRevLett.83.3538,
PhysRevLett.89.247003, PhysRevB.73.155113} is an exotic quantum phase where a local current of itinerant electrons forms a closed path.
Since charge current breaks the time-reversal symmetry, the loop current can be treated as an order parameter, which describes the breaking of this symmetry.
Such loop-current phase has been proposed in the studies of high-$T_{\rm c}$ cuprates, \cite{PhysRevB.37.3774, PhysRevB.55.14554, PhysRevLett.83.3538, PhysRevLett.89.247003, PhysRevB.73.155113} where the Coulomb interaction is expected to play an important role in the emergence of various phases.
For instance, Varma has studied a three-band Hubbard model for the cuprates, and found these loop-current phases stable in some parameter regions\cite{PhysRevB.55.14554, PhysRevLett.83.3538, PhysRevLett.89.247003, PhysRevB.73.155113}.
Furthermore, a recent numerical study by the variational cluster approach (VCA) has shown that the loop-current phase is a metastable
phase in the single-band Hubbard model on a square lattice.\cite{PhysRevB.85.125117}

\begin{figure}
\begin{center}
\includegraphics[scale=0.4]{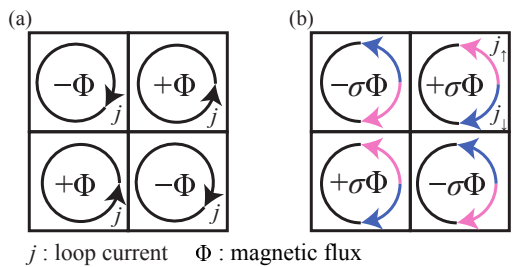}
\caption{(Color online) (a) Schematic diagram of the loop-current phase and the accompanying magnetic flux. Here, $\Phi$ is the magnitude of the magnetic flux and arrows indicate the direction of the current $j$. (b) Schematic diagram of the loop-spin-current phase. Here, pink (blue) arrows indicate the directions of the loop current from the up (down) spin $j_{\uparrow}$ ($j_{\downarrow}$) and $\sigma $ is $+1 (-1)$ for the up (down) spin.}
\label{intro}
\end{center}
\end{figure}

These theoretical studies have shown the existence of the loop-current phase induced by the Coulomb interaction.
This loop-current order can be regarded as a magnetic flux as depicted in Fig.~\ref{intro}(a), which is inferred from an intuitive picture based on the classical electrodynamics. 
Thus, the electronic states of the loop-current phase are similar to those of a system with a magnetic flux.
In other words, the Coulomb interaction can induce an effective magnetic flux, which breaks the time-reversal symmetry of the system.

In contrast, we can define another current phase which preserves the time-reversal symmetry by virtue of the electron spin degrees of freedom.
If the loop-current of up electron $j_{\uparrow}$ and that of down electron $j_{\downarrow}$ have opposite directions, $j_{\uparrow} = -j_{\downarrow}$,  as depicted in Fig.~\ref{intro}(b), a net loop-current, $j_{\uparrow}+j_{\downarrow}$, cancels out;
only the difference between the loop currents, $j_{\uparrow}-j_{\downarrow}$, has finite value. This difference corresponds to the cyclic flow of the electron spin, which we call the {\it loop-spin current}. This loop-spin current does not violate the time-reversal symmetry, but violates the rotational symmetry in spin space.

In terms of a magnetic flux, this loop-spin-current phase can be regarded as the states with a  {\it spin-dependent flux}: a magnetic flux that 
acts oppositely for opposite spin as denoted symbolically by $\sigma \Phi$  in Fig.~\ref{intro}(b).
Thus, the electronic states of loop-spin-current phase are essentially identical to those of the system with the spin-dependent flux.
Moreover, a theoretical study based on an extended Hubbard model \cite{PhysRevLett.100.156401} has shown that the Coulomb interaction induces this 
loop-spin-current phase.
Consequently, the Coulomb interaction can give rise to an effective spin-dependent flux.

The states with the magnetic flux or the spin-dependent flux may have close relation to topologically nontrivial states,\cite{RevModPhys.82.3045,
RevModPhys.83.1057} which have recently attracted considerable interest in condensed matter physics.
For example, the Haldane model,\cite{PhysRevLett.61.2015} which possesses a local magnetic flux, has the nontrivial Thouless-Kohmoto-Nightingale-den Nijs (TKNN) invariants.\cite{PhysRevLett.49.405}
This nontrivial TKNN invariants guarantee the quantization of the Hall conductance and the existence of the edge states. 
Moreover, the states with the nontrivial TKNN invariants are topologically protected.
In other words, the TKNN invariants cannot change without closing a band gap.\cite{PhysRevLett.49.405}

On the other hand, the system with the spin-dependent flux may have other topological invariants called the $Z_2$ invariants,\cite{PhysRevLett.95.146802} 
which ensure the existence of the time-reversal symmetric edge states.
The Kane-Mele model\cite{PhysRevLett.95.146802, PhysRevLett.95.226801} possesses the nontrivial $Z_2$ invariant and terms representing an intrinsic spin-orbit interaction, which causes topologically nontrivial characters.
Essentially, the effect of the spin-orbit interaction in this model is equivalent to that of the spin-dependent flux for itinerant electrons.

Consequently, these states with the magnetic flux or the spin-dependent flux may have the nontrivial topological invariants.
Furthermore, as shown in previous studies,\cite{PhysRevB.55.14554, PhysRevLett.83.3538, PhysRevLett.89.247003, PhysRevB.73.155113,PhysRevB.85.125117,PhysRevLett.100.156401} the Coulomb interaction may induce the loop-current (loop-spin-current) phase, whose electronic states are similar to those of the system with the magnetic (spin-dependent) flux.
In short, these loop-current and loop-spin-current phases may be considered as topological Mott insulating (TMI) phases: \cite{PhysRevLett.100.156401, PhysRevB.81.085105, PhysRevB.82.075125, JPSJ.80.044708, PhysRevB.84.201103} topologically nontrivial phases induced by the Coulomb interaction.
It should be noted that the TMI phases in this context are not necessarily Mott insulators.
In these phases, the word ``Mott'' only means ``induced by interactions''.

The previous mean-field analyses of the TMI phases have shown that a strong inter-site Coulomb interaction is required for stabilizing these phases 
in the single-band Hubbard model,\cite{PhysRevLett.100.156401, PhysRevB.81.085105, PhysRevB.82.075125, JPSJ.80.044708} since current orders are extracted from the mean-field decoupling of this inter-site interaction. 
However, as described earlier, it has been shown that the loop-current phase on a square lattice \cite{PhysRevB.85.125117} is metastable; the free energy of the loop-current phase is lower than that of the trivial phase, and higher than that of the antiferromagnetic phase. 
In contrast, on a half-filled honeycomb lattice, strong on-site $U$ is required for the emergence of the antiferromagnetic phase because of the semimetallic behavior.\cite{PhysRevLett.110.096402}
Considering this fact, the simple Hubbard model on a honeycomb lattice may favor the TMI phases induced by the loop-current or loop-spin-current order.

In this paper, we explore the possibility of the TMI phases induced by the loop-current or loop-spin-current order 
in the simple Hubbard model on a honeycomb lattice.
The VCA takes into account appropriately the effects of strong short-range correlations due to the on-site Coulomb interaction.
Moreover, it turns out that our theory does not require the inter site Coulomb interaction to describe the loop-current or loop-spin-current order.
Thanks to these advantages, the VCA seems to be a suitable method for our purpose.
As a result of the VCA calculation, we obtain a stable loop-spin-current phase which has not been reported before. 
We also find that this phase has a very small but finite single-particle gap and is characterized by the nontrivial $Z_2$ invariants.

\section{Methods \label{methods}} 
In order to investigate the existence of the loop-current or loop-spin-current phase, we apply the VCA to the single-band Hubbard model on a honeycomb lattice.
The Hamiltonian $H$ is given by 
\begin{equation}
H = -t \sum_{\langle ij \rangle \sigma} (c^{\dagger}_{i \sigma} c_{j \sigma} + {\rm H.c.}) + U \sum_{i} n_{i \uparrow} n_{i \downarrow} - \mu \sum_{i \sigma} n_{i \sigma} ,
\end{equation}
where $\langle ij  \rangle$ denotes nearest-neighbor pairs, $t$ is a hopping integral, $c_{i \sigma}$ is the annihilation operator of an electron at site $i$ with
spin $\sigma$, $n_{i \sigma} = c^{\dagger}_{i \sigma} c_{i \sigma}$, $\mu$ is a chemical potential, and $U$ is the on-site Coulomb repulsive interaction.
\begin{figure}
\includegraphics[scale=0.4]{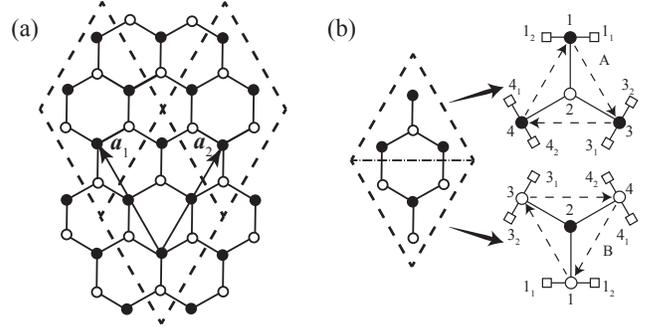}
\caption{(a) A cluster tiling (dashed lines) on a honeycomb lattice used for calculations. A superlattice consists of the same eight-site cluster.
Symbols $\circ$ and $\bullet$ denote two inequivalent sites, $\bm{a}_1$ and $\bm{a}_2$ represent the superlattice unit vectors. (b) The decomposition of the eight-site cluster into two four-site subclusters A and B. Each subcluster has six-site bath sites.
A symbol $\Box$ denotes the bath site attached to the cluster site, and dashed arrows denote the loop-current and loop-spin-current order.
}
\label{cluster}
\end{figure}
In the VCA, the lattice is tiled into the set of the same clusters as shown in Fig.~\ref{cluster}(a). 
Hereafter, we call this set of the clusters the {\it superlattice}.
Within each cluster, we define the cluster Hamiltonian $H^{\prime}$, which does not contain terms connecting different clusters.
In reciprocal space, a wavenumber $\bm{k}$ is written as $\bm{k} = m \bm{b}_1 + n \bm{b}_2$, where $\bm{b}_i$
 is the reciprocal vector of $\bm{a}_i$ in Fig.~\ref{cluster}(a). 
In this representation, the Brillouin zone is defined by $-\frac{1}{2} \leq m \leq \frac{1}{2} $ and $-\frac{1}{2} \leq n \leq \frac{1}{2}$.

The VCA is based on the self-energy-functional theory,\cite{potthoff2003self} in which we solve a variational problem for the Potthoff
functional defined as
\begin{equation}
\label{eq:Legendre}
\Omega [\Sigma] = \mathrm{Tr} \ln [-({\bm G}_{0}^{-1} - \Sigma)^{-1}] + F[\Sigma].
\end{equation}
Here, $\Sigma$ is the self-energy of the system, ${\bm G}_{0}$ is the free Green's function, $F[\Sigma] = \Phi[{\bm G}]-\mathrm{Tr}(\Sigma {\bm G})$ 
is the Legendre transform of the Luttinger-Ward functional $\Phi[{\bm G}]$,\cite{PhysRev.118.1417} where ${\bm G}$ is the 
Green's function of the system.
Since the Green's function is given by the equality, 
${\bm G} = -\frac{\delta F[\Sigma]}{\delta \Sigma}$,\cite{potthoff2003self} the variational condition of Eq.~(\ref{eq:Legendre}) 
is represented as
\begin{equation}
\label{eq:Dyson}
\frac{\delta \Omega [\Sigma]}{\delta \Sigma} =  -{\bm G} + ({\bm G}_{0}^{-1} -\Sigma)^{-1} = 0.
\end{equation}
This condition is equivalent to the Dyson equation,
${\bm G}^{-1} = {\bm G}_{0}^{-1} - \Sigma$.
Furthermore, using the solution of Eq.~(\ref{eq:Dyson}), $\Sigma_{\mathrm{sol}}$, an exact thermodynamic potential is given by $\Omega[\Sigma_{\mathrm{sol}}].$
\cite{potthoff2003self}

The exact self-energy $\Sigma_{\mathrm{sol}}$ can be obtained by solving the original Hamiltonian $H$.
However, it is almost impossible to solve $H$ exactly.
Instead, we prepare a trial self-energy, i.e., adequate substitutes for $\Sigma_{\mathrm{sol}}$.
In strongly correlated systems, where the local Coulomb interaction plays an important role, the self-energy including short-range correlations is expected to be 
an appropriate trial self-energy.
Thus, in the VCA, we adopt the trial self-energy $\Sigma^{\prime}$ which is derived from the small cluster Hamiltonian $H^\prime$ with the local interaction terms.
If we equate interaction terms of $H^\prime$ to those of the original Hamiltonian $H$, the Potthoff functional $\Omega[\Sigma^{\prime}]$
is represented as \cite{potthoff2003self}
\begin{equation}
\label{eq:potthoff}
\Omega [\Sigma^{\prime}] = \Omega^\prime + \mathrm{Tr} \ln [ -({\bm G}^{-1}_{0}-\Sigma^{\prime})^{-1} ] 
- \mathrm{Tr} \ln(-{\bm G}^\prime) .
\end{equation}
Here, $\Omega^{\prime}$ is the thermodynamic potential of the cluster and $\bm{G}^\prime$ is the exact Green's function of $H^\prime$.
 
Unlike the interaction terms, we can add any one-body terms to $H^{\prime}$ in the VCA. 
These degrees of freedom for one-body terms give
variety to the trial self-energy. 
Then, we treat one-body parameters ${\bm t}^\prime$ in the cluster as variational parameters for the trial
self-energy, i.e., $\Sigma^\prime = \Sigma^{\prime}({\bm t}^\prime)$.
Practically, the variational condition is represented in terms of variational parameters as follows:
\begin{equation}
\label{eq:vp}
\frac{\delta \Omega[\Sigma^{\prime}({\bm t}^\prime)]}{\delta {\bm t}^\prime}  = 0.
\end{equation}

The choice of the cluster and its one-body terms restricts the functional space of the trial self-energy.
Equation (\ref{eq:vp}) represents the variational condition for the self-energy-functional in the restricted functional space.
This restriction makes the VCA approximative.

In the VCA, symmetry breakings are described by fictitious external fields called the Weiss fields.\cite{PhysRevB.70.245110}
We consider a cluster whose symmetry is lowered by the Weiss fields, and treat these amplitudes as variational parameters.
A symmetry-breaking state is characterized by the solution of Eq.~(\ref{eq:vp}) with finite amplitudes of the Weiss fields.

Imitating the mean-field analyses,\cite{PhysRevLett.100.156401,PhysRevB.81.085105} we adopt next-nearest-neighbor hopping terms as the Weiss-field terms for the TMI phases.
Specifically, these Weiss-field terms are defined as 
\begin{equation}
\label{eq:qah}
H_{\rm W}^{a} = {\rm i} \lambda_{a} \sum_{\langle \langle ij \rangle \rangle \sigma} (\nu_{ij} c^{\dagger}_{i \sigma} c_{j \sigma}
+{\rm H.c.})
\end{equation}
for the loop-current phase, and
\begin{equation}
H_{\rm W}^{s} = {\rm i} \lambda_{s} \sum_{\langle \langle ij \rangle \rangle \alpha \beta} 
\sigma^z_{\alpha \beta} (\nu_{ij} c^{\dagger}_{i \alpha} c_{j \beta}
+{\rm H.c.})
\label{eq:qsh}
\end{equation}
for the loop-spin-current phase.
Here, $\langle \langle ij \rangle \rangle$ denotes next-nearest-neighbor pairs, $\lambda_{x}$ ($x = a,s$) is the amplitude of the Weiss fields, $\sigma^z$ is the $z$-component of the Pauli matrices, and an antisymmetric matrix $\nu_{ij}$ represents how the electron circulates;
if the electron turns left (right) in next-nearest-neighbor hopping from site $j$ to $i$, $\nu_{ij} = -\nu_{ji}=+1(-1)$.
The Weiss field $\lambda_{a}$ $(\lambda_{s})$ describes the spontaneous loop-current (loop-spin-current) order depicted in Fig.~\ref{cluster}(b). The sign of the Weiss field represents the direction of the spontaneous current; the positive Weiss field corresponds to the current along the direction of the arrow. 
We note that the corresponding topological invariants must be evaluated, because it is not obvious that these current ordered phases have nontrivial topological numbers.

Following a recent study within the VCA on a honeycomb lattice,\cite{PhysRevLett.110.096402} we divide the eight-site cluster into two four-site clusters with six-site bath
sites as illustrated in Fig.~\ref{cluster}(b).
We call these two four-site clusters the subcluster A and B, respectively. The trial self-energy is obtained by solving two subcluster problems independently.
The cluster we adopt can describe the electronic states on a honeycomb lattice most appropriately within the present VCA.\cite{PhysRevLett.110.096402}

In this paper, from the above arguments, the cluster Hamiltonian $H^{\prime}$ is given by the direct sum of 
$H_{\rm sub}^A$ and $H_{\rm sub}^B$ , where $H_{\rm sub}^\eta$ is defined within the subcluster $\eta$ ($\eta = {\rm A, B}$) as
\begin{eqnarray}
H_{\rm sub}^{\eta} &= -t {\displaystyle\sum_{\langle ij \rangle \sigma}} (c^{\dagger}_{i \sigma} c_{j \sigma} + {\rm H.c.}) + U {\displaystyle \sum_{i}} n_{i \uparrow} n_{i \downarrow} - \mu {\displaystyle \sum_{i \sigma}} n_{i \sigma} \nonumber \\
&+  {\displaystyle \sum_{il \sigma}}( \theta_{i_l} c^{\dagger}_{i \sigma}a_{i_l \sigma} +{\rm H.c.} )
+ {\displaystyle \sum_{i l \sigma}} \epsilon_{i_ l} a^{\dagger}_{i_l \sigma}a_{i_l \sigma} + H_{\rm W}^{x}.
\label{eq:subHam}
\end{eqnarray}
Here, $a_{i_l \sigma}$ is the annihilation operator of an electron with spin $\sigma$ at $l$-th ($l = 1,2$) bath site attached to $i$-th cluster site, $\theta_{i_l}$
is a ``hopping'' parameter between $i$-th cluster site and $l$-th bath site, $\epsilon_{i_l}$ is an on-site energy of $l$-th bath site attached to $i$-th cluster site. 

We consider only the half-filling case where the particle-hole symmetry and the
cluster symmetry restrict the one-body parameters as $\mu = \frac{U}{2}$,
$\theta_{i_l} = \theta$, and $\epsilon_{i_l} = (-1)^l \epsilon$.\cite{PhysRevLett.110.096402}
Then, three one-body parameters $\theta$, $\epsilon$, and $\lambda_{x} $ are treated as variational parameters for the optimization of the self-energy.
The Potthoff functional also depends on variational parameters $\theta$, $\epsilon$, and $\lambda_{x} $. For simplicity,
we introduce the notation $\Omega(\lambda_x) \equiv \Omega[\Sigma^{\prime}(\lambda_x, \theta_{\mathrm{opt}}, \epsilon_{\mathrm{opt}})]$,
where $\theta_{\mathrm{opt}}$ and $\epsilon_{\mathrm{opt}}$ are defined by the simultaneous conditions, $\left. \frac{\partial \Omega(\lambda_x, \theta, \epsilon_{\mathrm{opt}})}{\partial \theta} \right |_{\theta=\theta_{\mathrm{opt}}}= 0$ and $ \left. \frac{\partial \Omega(\lambda_x, \theta_{\mathrm{opt}}, \epsilon)}{\partial \epsilon} \right |_{\epsilon=\epsilon_{\mathrm{opt}}} = 0$. 
In order to obtain the trial self-energy, we use the band Lanczos algorithm.\cite{bai1987templates}
Since we consider only the system with moderate $U$ ($U \leq 3.5\,t$) in this paper, the antiferromagnetic phase, which exists for larger $U$ ($U \gtrsim 4\,t$),
\cite{PhysRevLett.110.096402} is not taken into consideration.

By using the optimized variational parameters, the single-particle Green's function of the system $\bm{G}$ is given by
\begin{equation}
\bm{G}=\frac{1}{\bm{G}^{-1}_{0}-\Sigma^{\prime}({\bm t}^\prime)} .
\end{equation}
Physical quantities of the system are calculated from this Green's function $\bm{G}$. For instance, the magnitude of the loop-spin-current $j_s$ is given by
\begin{equation}
j_s = -{\rm i} t \sum_{\alpha \beta} \sigma^z_{\alpha \beta} \left( \langle c^{\dagger}_{i \alpha} c_{j \beta} \rangle - \langle
c^{\dagger}_{j \beta}c_{i \alpha} \rangle \right) ,
\end{equation}
where $i$ denotes the site corresponding to the starting point of the arrow, $j$ denotes the site corresponding to the point of the arrow in Fig.~\ref{cluster}(b), and bracket denotes the expectation value of an one-body operator, which is calculated from the Green's function as follows:
\begin{equation}
\langle  c^{\dagger}_{\beta} c_{\alpha} \rangle = \sum_{\bm{k}} \int_{C} \frac{{\rm d} z}{2 \pi {\rm i}} G_{\alpha \beta}(z,\bm{k}).
\end{equation}
Here, a contour $C$ surrounds the negative real frequency axis counterclockwise.

The topological numbers of interacting systems are determined via the single-particle Green's function $\bm{G}$. \cite{PhysRevLett.105.256803,PhysRevX.2.031008}
It is very useful to introduce the topological Hamiltonian \cite{PhysRevX.2.031008} when calculating the topological numbers of interacting systems.
From the zero-frequency value of the Green's function $\bm{G}$, the topological Hamiltonian $H_{\rm top}(\bm{k})$ is defined by
\begin{equation}
H_{\rm top}(\bm{k}) = -\bm{G}^{-1}(0,\bm{k}).
\end{equation}

In the topological Hamiltonian formalism, the topological numbers of {\it interacting systems} are given by the similar expressions of the topological numbers defined in {\it non-interacting systems}.\cite{PhysRevX.2.031008}
For non-interacting two-dimensional systems, the topological numbers are defined by the occupied eigenstates of Hamiltonian in $k$ space $H_0 (\bm{k})$.
The TKNN invariants $C_1$ \cite{PhysRevLett.49.405} are defined as
\begin{equation}
\label{eq:tknn}
C_1 = \frac{1}{2 \pi} \int {\rm{d} }^2 \bm{k} f_{xy}. 
\end{equation}
Here, the integration is over the Brillouin zone, $f_{ij} = \partial_i a_j -\partial_j a_i$, and $a_i = - {\rm i}  \sum_{\alpha} \braket{u^{\alpha}(\bm{k}) | \partial_{k_i} |u^{\alpha}(\bm{k})}$, where $\ket{u^{\alpha} (\bm{k})}$ is the $\alpha$-th eigenstate of $H_0 (\bm{k})$
 and $\alpha$ runs through all the occupied bands.
The $Z_2$ invariants $\Delta$ \cite{PhysRevB.74.195312} are defined as
\begin{equation}
\label{eq:z2}
(-1)^{\Delta} = \prod_{\Gamma_i = {\rm TRIM} } \frac{\sqrt{\mathrm{det} \bm{B}(\Gamma_i)} }{\mathrm{Pf}(\bm{B}(\Gamma_i))}, 
\end{equation}
where ``TRIM'' stands for four time-reversal invariant momenta: $\frac{n_1}{2}\bm{b}_1 + \frac{n_2}{2}\bm{b}_2 $ with $ n_1, n_2 = 0,1$. 
Here, the matrix $\bm{B(k)}$ is defined by $B_{\alpha
\beta}(k) = \braket{u^{\alpha}(-\bm{k}) | \hat{T} | u^{\beta}(\bm{k})}$,
and $\hat{T}$ is the time-reversal operator. 
If we replace $H_0 (\bm{k})$ by $H_{\rm top} (\bm{k})$, Eqs.~(\ref{eq:tknn}) and (\ref{eq:z2}) give the topological invariants for interacting systems.

In general, the numerical evaluation of the topological numbers is difficult because of its gauge dependence.\cite{PhysRevB.83.235115}
Thus, some gauge-invariant methods have been proposed for calculating the topological numbers numerically.\cite{PhysRevB.83.235115, JPSJ.74.1674,PhysRevB.84.075119}
With the topological Hamiltonian formalism, we can use every gauge-invariant method for non-interacting systems in order to evaluate the topological invariants for interacting systems.
In this paper, the gauge-invariant methods\cite{JPSJ.74.1674,PhysRevB.84.075119} that have been proposed for non-interacting topological invariants are adopted.

\section{Results \label{results}}
\begin{figure}
\includegraphics[scale = 0.7]{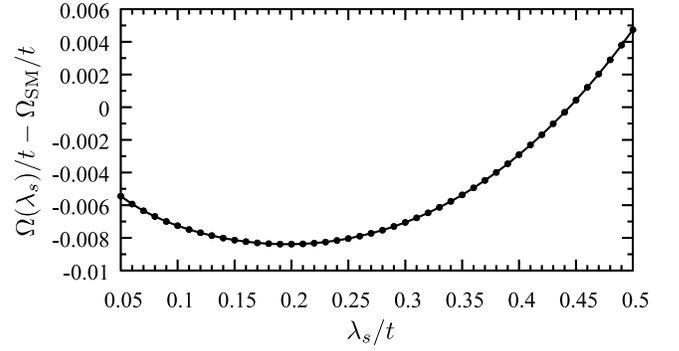}
\caption{The Potthoff functional $\Omega$ in the loop-spin-current case at $U = 3.5\,t$.  The thermodynamic potential of the semimetal phase 
$\Omega_{\mathrm{SM}}$ is obtained by the optimization of the functional without the-Weiss field terms.}
\label{fig:lsp}
\end{figure}

In the VCA, the existence of ordered phases can be determined by whether the Potthoff functional $\Omega(\lambda_x)$ has a stationary point with a nonzero Weiss field $\lambda_x \neq 0$. At the stationary point, the value of the Potthoff functional is equivalent to the thermodynamic potential of the ordered phase as explained in the previous section.
Figure \ref{fig:lsp} represents the Potthoff functional $\Omega(\lambda_{s})$ in the loop-spin-current case at $U=3.5\,t$.
The functional $\Omega(\lambda_s)$ has a minimum around $\lambda_s \approx 0.2\,t$, which indicates the existence of the loop-spin-current phase.
We have verified that the thermodynamic potential of the loop-spin-current phase is lower than that of the semimetal phase.
These facts show clearly that the loop-spin-current phase is energetically stable within the VCA.

\begin{figure}
\includegraphics[scale = 0.7]{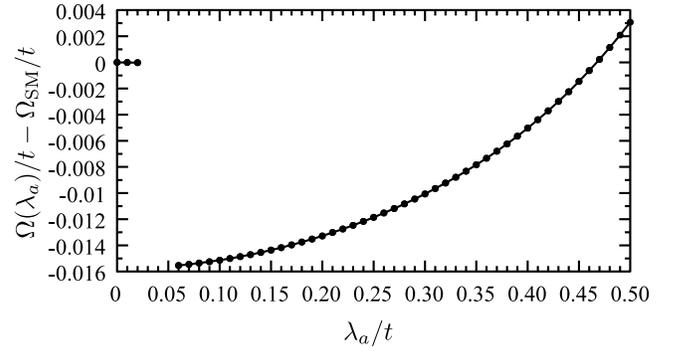}
\caption{The Potthoff functional $\Omega$ including the loop-current-Weiss field $\lambda_{a}$ at $U = 3.5\,t$. The discontinuity exists around $\lambda_{a} = 0.06t$. For the region $0.03t \leq \lambda_{a} < 0.06t$, we cannot find a stationary point of the Potthoff functional.}
\label{fig:loop}
\end{figure}

Figure \ref{fig:loop} shows the Potthoff functional $\Omega(\lambda_{a})$ in the loop-current case at $U=3.5\,t$. In contrast to the loop-spin-current case, the Potthoff functional $\Omega(\lambda_a)$
 has the discontinuity near the stationary point $\lambda_a \approx 0.06\, t$. Such discontinuity has been reported in the previous VCA study by Potthoff,\cite{potthoff2003self} who has pointed out that a discontinuity near a stationary point is an artifact due to the choice of a reference system.
The discontinuity of the present functional $\Omega(\lambda_a)$ may indicate that the reference system we choose is insufficient for describing the loop-current phase.
Since the discontinuity exists only when $\lambda_a$ is finite, we guess that the cause of the discontinuity exists in the Weiss-field terms.
Consequently, a modification of the Weiss-field terms in the loop-current case is attempted in order to remove the discontinuity.
The Weiss-field terms defined by Eq.~(\ref{eq:qah}) cannot describe inter-cluster-current orders, though the Haldane model contains all the next-nearest-neighbor hopping terms with imaginary hopping integral.
 
\begin{figure}
\includegraphics[scale=0.7]{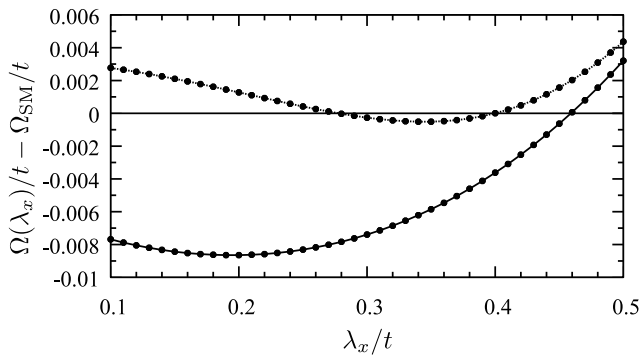}
\caption{The Potthoff functional $\Omega$ including the Weiss fields $\tilde{\lambda}_{x} \ (x = s,a)$ for the loop-spin-current case (solid line) and the loop-current case (doted
line) at $U = 3.5\,t$. For each $\lambda_{x}$, the bath parameters, $\theta$, $\epsilon$, and $\tilde{\lambda}_{x}$ are optimized.}
\label{gp}
\end{figure}

Hereafter, we take into account effectively the inter cluster-current order. 
The bath sites in the reference system represent the environment system around the cluster. 
We define the Weiss-field terms describing the loop-current order between the cluster and the bath sites
as follows:
\begin{eqnarray}
\label{eq:bathqah}
H_{{\rm W} b}^{a} = {\rm i} \tilde{\lambda}_{a} \sum_{l \sigma} \sum_{i \neq 2}(-1)^l c^{\dagger}_{2 \sigma} a_{i_l \sigma} +{\rm H.c.},
\end{eqnarray}
where $\tilde{\lambda}_{a}$ is the amplitude of the Weiss fields [See Fig.~\ref{cluster}(b) for the definition of site indices].
Since Eq. (\ref{eq:bathqah}) connects the environment system and the cluster site,
this term can be thought as a substitute for the inter-cluster-current terms.
We also define the Weiss-field terms
\begin{equation}
H_{{\rm W}b}^{s} = {\rm i} \tilde{\lambda}_{s} \sum_{l \alpha \beta} \sum_{i \neq 2} 
\sigma^z_{\alpha \beta} (-1)^l c^{\dagger}_{2 \alpha} a_{i_l \beta}
+{\rm H.c.}
\label{eq:bathqsh}
\end{equation}
for the loop-spin-current phase.
Using the subcluster Hamiltonian given by the summation of Eq. (\ref{eq:subHam}) and the inter-loop-current terms (\ref{eq:bathqah}) or (\ref{eq:bathqsh}) , we reevaluate the Potthoff functionals. 
Figure \ref{gp} shows the reevaluated Potthoff functionals $\Omega(\lambda_a)$ and $\Omega(\lambda_s)$, which do not have any discontinuities.
The reevaluated Potthoff functional $\Omega(\lambda_a)$ [$\Omega(\lambda_s)$] has a minimum, which corresponds to the loop-current (loop-spin-current) phase.
We have verified that the thermodynamic potentials of both phases are lower than that of the semimetal phase.
Moreover, as seen in Fig.~\ref{gp}, the loop-spin-current phase is energetically more stable than the loop-current phase.
Thus, the loop-spin-current phase is the most energetically stable phase in this analysis.

\begin{figure}
\includegraphics[scale=0.71]{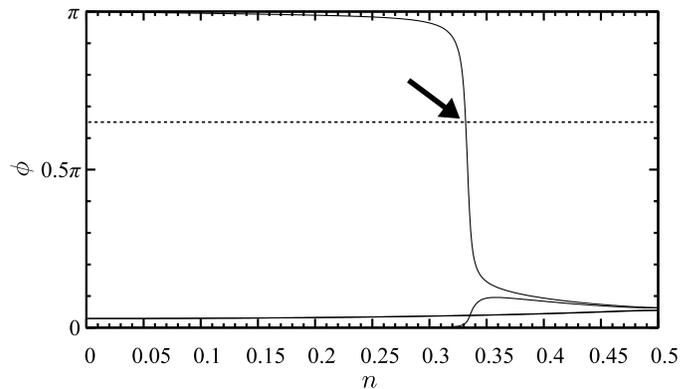}
\caption{The evolution of the Wannier function center $\phi$ (solid lines) at $U = 3.5\,t$. Since $\phi$ are given by phases of eigenvalues of the
matrix,\cite{PhysRevB.84.075119} this value is limited in the region $- \pi \leq \phi \leq \pi $. The arrow indicates the cross point of the $\phi$ curve and the reference line (dotted lines).
Since the obtained $\phi$ curves are symmetric with respect to the $n$ axis, we omit the negative $\phi$ region.}
\label{wannier}
\end{figure}

In the loop-spin-current phase, we evaluate the $Z_2$ invariants from the evolution of the Wannier function centers\cite{PhysRevB.84.075119} of the topological
Hamiltonian $H_{\rm top}$.
Figure \ref{wannier} represents the behavior of each Wannier function center $\phi$ at $U=3.5\,t$.
Here, the wavenumber parameter $n$ in the horizontal axis corresponds to the wavenumber parallel to the reciprocal vector $\bm{b}_2$  (See Sec. \ref{methods}), and we integrate out the wavenumber parameter $m$.
The $Z_2$ invariants can be obtained from these curves by drawing an arbitrary line parallel to the $n$ axis, and counting how many times this line crosses $\phi$ curves.
If they cross odd number times, the $Z_2$ invariant of the system is nontrivial.
Since the line parallel to the $n$ axis crosses the $\phi$ curves odd number times, we find that this phase has nontrivial $Z_2$ invariants.
Therefore, this loop-spin-current-ordered phase is considered to be the quantum spin Hall (QSH) state, 
which possesses the nontrivial $Z_2$ invariant and satisfies the spin conservation law.

For the loop-current phase, we employ an efficient algorithm \cite{JPSJ.74.1674} to evaluate the TKNN invariants. As a result, the loop-current phase has
nonzero TKNN invariants $C_1 = 2$. Thus, this loop-current-ordered phase is considered to be the quantum anomalous Hall (QAH) state: the quantum Hall state
 without an external magnetic flux.
The TKNN invariants become even number because of the spin degrees of freedom.

Hereafter, we concentrate our attention on the QSH state, which is energetically more stable than the other phases.
In the QSH state, the $Z_2$ invariants are protected by the time-reversal symmetry and the existence of a single-particle gap.
Because of this nature, the QSH state exist until the single-particle gap closes.
In order to evaluate the single-particle gap $\Delta_{\rm sp}$, we calculate the difference between the lowest positive pole $\omega_{\rm e}$ 
and the highest negative pole $\omega_{\rm h}$, i.e., $\Delta_{\rm sp} = \omega_{\rm e} - \omega_{\rm h}$.\cite{PhysRevB.77.045133}
\begin{figure}
\includegraphics[scale=0.6]{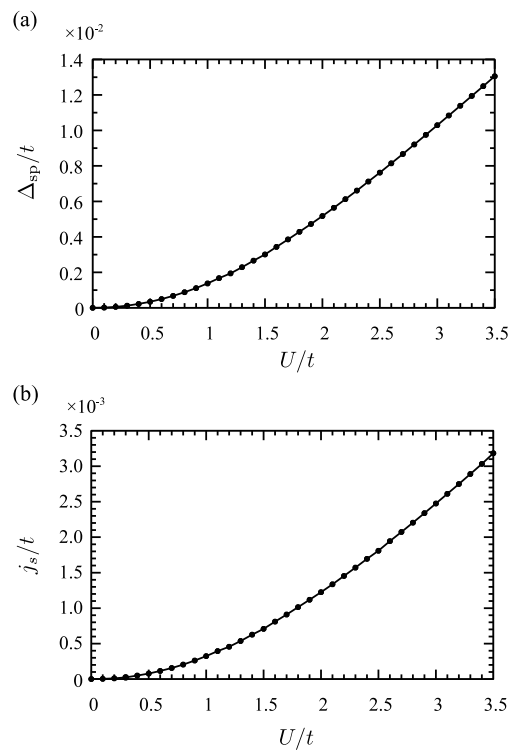}
\caption{(a) The single-particle gap $\Delta_{\rm sp}$ as a function of the on-site Coulomb interaction $U$. (b) The magnitude of the
loop-spin-current $j_s$ as a function of the on-site Coulomb interaction $U$. }
\label{gap}
\end{figure}

Figure \ref{gap}(a) gives the single-particle gap $\Delta_{\rm sp}$ as a function of the on-site Coulomb interaction $U$.
For a wide range of interaction parameters $U$, there exists the QSH state with very small but finite single-particle gap $\Delta_{\rm sp}$, whose size is about one-thousandth of the band width.
Following the same evaluation process for the $Z_2$ invariants represented above, we confirm the existence of the QSH state at each interaction parameter $U$.
Furthermore, the fact that the single-particle gap $\Delta_{\rm sp}$ is increasing with the Coulomb interaction $U$ suggests that the QSH state is induced by this on-site interaction.
Figure \ref{gap}(b) represents the $U$ dependence of the magnitude of the loop-spin-current $j_s$, which is very similar to Fig.~\ref{gap}(a).
This type of similarity has been reported in the previous studies; the band gap of the Kane-Mele model is proportional to the magnitude of the intrinsic spin-orbit coupling.\cite{PhysRevLett.95.146802,PhysRevLett.95.226801}
Thus, the single-particle gap in the QSH state is not the Mott one; this gap is considered to be induced by the spontaneous loop-spin current.

\section{Cluster Dependence}
\begin{figure}
\begin{center}
\includegraphics[scale=0.4]{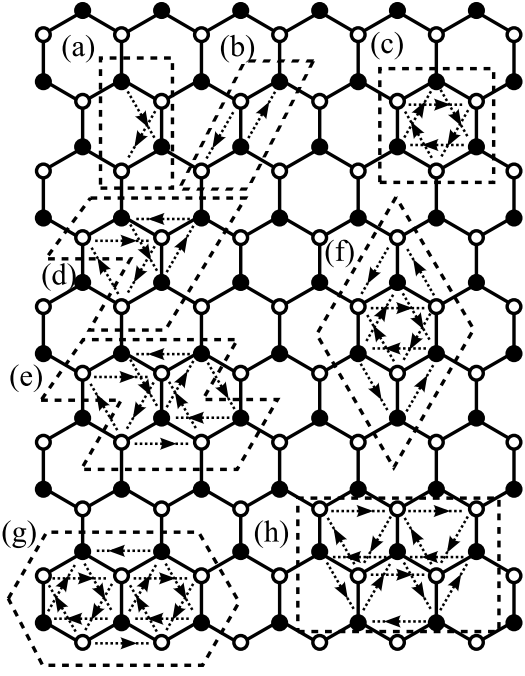}
\caption{Clusters used for studying cluster dependence. We prepare eight clusters: cluster (a)-(h).
Dotted arrows in the cluster denote the loop-spin-current order. \label{fig:reference}}
\end{center}
\end{figure}

Since the results in the previous section are given by the calculation based on a single cluster,
it is possible that the energetically stable loop-spin-current phase is peculiar to certain cluster choice.
In order to eliminate this possibility,
we examine the existence of the loop-spin-current phases in other eight clusters shown in Fig. \ref{fig:reference}.
For simplicity, we consider only the loop-spin-current phase at $U=3.5\, t$,
and adopt a simple cluster Hamiltonian as follows:
\begin{eqnarray}
H_{\mathrm{cl}} = &-&t^{\prime} \sum_{\langle ij \rangle \sigma}(c^{\dagger}_{i \sigma} c_{j \sigma}
 + \mathrm{H.c.}) + \mathrm{i} \lambda
\sum_{\langle \langle ij \rangle \rangle \alpha \beta}
\sigma^z_{\alpha \beta}(\nu_{ij} c^{\dagger}_{i \alpha} c_{j \beta} + \mathrm{H.c.}) \nonumber \\
&+& U\sum_{i} n_{i \uparrow}
n_{i \downarrow} - \mu \sum_{i \sigma}n_{i \sigma}.
\end{eqnarray}

Since the particle-hole symmetry restricts the chemical potential $\mu = \frac{U}{2}$ at half-filling,
one-body parameters $t^{\prime}$ and $\lambda$ are treated as the variational parameters.
It should be noted that these clusters may have poor ability to describe the semimetal phase\cite{PhysRevLett.110.096402} and results based on these clusters are not very reliable quantitatively.
However, we consider that these clusters are sufficient for examining tendencies towards the loop-spin-current phases.

\begin{figure}
\begin{center}
\includegraphics[scale=0.32]{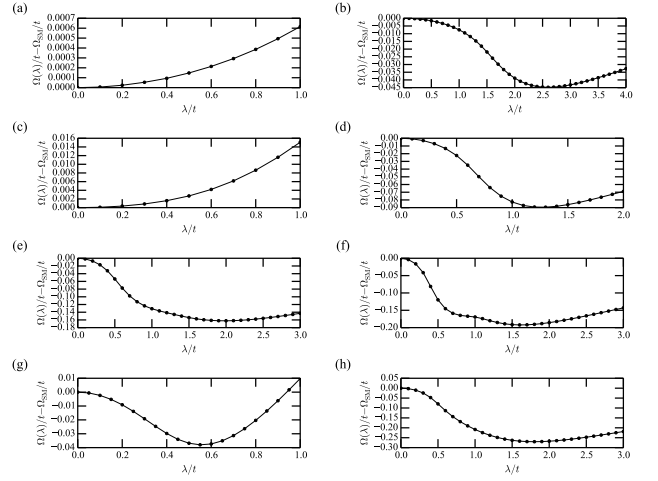}
\caption{The Potthoff functionals including loop-spin-current order calculated with clusters shown€ in Fig. \ref{fig:reference}. For each $\lambda$, we optimize the parameter $t^{\prime}$. We do not divide the values of the Potthoff functional by the size of cluster. The variation in the value of the Potthoff functional can be considered as a result of the strong dependence on cluster choices for the ability to describe the semimetal phase.}
\label{fig:clusterDepend}
\end{center}
\end{figure}

Figures \ref{fig:clusterDepend}(a)-\ref{fig:clusterDepend}(h) show the calculated Potthoff functionals $\Omega (\lambda)$ for each cluster.
As seen in Figs. \ref{fig:clusterDepend}(a)-\ref{fig:clusterDepend}(h), only the cluster (a) and cluster (c) do not indicate the tendency towards the loop-spin-current phase,
while the others show the existence of energetically stable loop-spin-current phases.
In order to understand this difference,
we conjecture a condition that a cluster should satisfy for the energetically stable loop-spin-current phase.
A current we assume belongs to a hexagonal plaquette.
Comparing the most simple clusters, cluster (a) and cluster (b),
we notice that two currents in cluster (a) belong to the same plaquette while those in cluster (b) belong to different plaquettes.
In cluster (c), every current also belongs to the same plaquette as shown in Fig. \ref{fig:reference}, 
and the loop-spin-current phase is not stable in this cluster choice.
From the above facts, we reach the conjecture that
{\it the loop-spin-current phase emerges if a cluster contains currents belonging to different hexagonal plaquettes.}
The results shown in Figs. \ref{fig:clusterDepend}(a)-\ref{fig:clusterDepend}(h) support our conjecture.

There exist only two clusters where every current order belongs to the same plaquette,
and clusters that consist of more than six sites always contain current orders belonging to different plaquettes.
Therefore, we conclude that the loop-spin-current phase in this analysis is not peculiar to a certain cluster choice.

\section{Discussion and Summary \label{discussion}}
The main goal of this analysis is to investigate the possibility of a topologically nontrivial state induced by the on-site Coulomb interaction.
As shown in the previous section, the results obtained by the VCA show clearly the existence of an energetically stable QSH state and an unstable QAH state induced by this interaction.
The appearance of this difference between these states is consistent with results of the previous study on TMI phases.\cite{PhysRevLett.100.156401}
According to the study based on the extended Hubbard model,\cite{PhysRevLett.100.156401} 
the QSH and the QAH states are degenerate within the mean-field approximation. 
When the effect of quantum fluctuation is considered, the difference of broken symmetries lifts this degeneracy.\cite{PhysRevLett.100.156401}
In the QAH state, the discrete time-reversal symmetry is broken, and no corresponding Nambu-Goldstone mode exists.
On the other hand, in the QSH state, continuous rotational symmetry in spin space is broken, and there exist Nambu-Goldstone modes.
Consequently, the existence of the Nambu-Goldstone mode affects the thermodynamic potential.
The difference of the thermodynamic potential
is a result of quantum fluctuations,
which the VCA is able to capture, while a mean-field approximation is not.
Therefore, our results are consistent with the previous study,\cite{PhysRevLett.100.156401} and this is the reason why we concentrate our attention only on the QSH state.

On the other hand, our results are inconsistent with quantum Monte Carlo simulations.
In the half-filled Hubbard model on a honeycomb lattice, quantum Monte Carlo simulations have been performed in order to investigate the existence of the quantum spin liquid state.\cite{nature08942, srep00992}
In the largest simulation,\cite{srep00992} it has been reported that there is no paramagnetic phase with finite single-particle gap.

We consider that this difference may reflect the limitation of our calculation.
Although the VCA describes effects from correlations within the cluster exactly, this approach may not be able to treat long-range correlations appropriately.
In a small scale of the cluster, it is highly probable that a spontaneous loop-spin current exists.
On the other hand, it may be hard to describe the large scale behavior of this current.
The loop-spin current obtained by this approach might be restricted to a finite range.
If this short-range order cannot develop into a long-range order, the difference is naturally understood;
the calculation based on the VCA overestimates the stability of the loop-spin-current phase.

Since long-range quantum fluctuations may suppress the development of the loop-spin-current order,
it is desirable that the effects of long-range fluctuations are irrelevant in order to find stable TMI phases induced by the on-site interaction.
One candidate for such TMI phases is the Hubbard model on a kagome lattice, which has a larger coordinate number than that of a honeycomb lattice and 
a Dirac point at one-third filling.
We left the investigation of this possibility for future work.

Whether or not the short-range order develops into the long-range order, 
our results show clearly that the on-site Coulomb interaction can induce this loop-spin-current order.
In the system with the loop-spin current, the electronic states might have a similarity to those of the system with spin-orbit interaction.
Therefore we consider that the on-site interaction generates, at least locally, the effective spin-orbit interaction.

Such an effective spin-orbit interaction induced by correlation effects has been proposed by Wu and Zhang.\cite{PhysRevLett.93.036403}
According to related studies,\cite{PhysRevLett.93.036403, PhysRevB.75.115103} 
this dynamically generated spin-orbit interaction is a result of Pomeranchuk instabilities, which come from the deformation of the Fermi surface. 
In the half-filled honeycomb system, however, this scenario cannot be applied directly, 
because the density of states vanishes at the Fermi level. In short, no Pomeranchuk instability occurs.

Consequently, the VCA calculation  implies the existence of an alternative mechanism for giving rise to the effective spin-orbit interaction dynamically.
Since our calculation is performed in the half-filled Hubbard model on a honeycomb lattice, 
the effective spin-orbit interaction from an alternative mechanism is not a result of Fermi-surface effects, and is induced by local interactions.
Therefore this alternative effective spin-orbit interaction is expected to be seen in many systems.
We also left the clarification of the mechanism that explains how the on-site interaction generates the effective spin-orbit interaction for future work.

In summary, we have shown the possibility of the topologically nontrivial states and the accompanying effective spin-orbit interaction induced by the on-site interaction.
Although this effective interaction may be suppressed by long-range fluctuations in the half-filled Hubbard model on a honeycomb lattice, 
we consider that a similar effective spin-orbit interaction can be seen in other systems, for instance, the one-third filled Hubbard model on a kagome lattice.
The existence of such an effective spin-orbit interaction generated from the simple on-site interaction leads to the ubiquity of topological Mott insulating phases.
Therefore, it is strongly required to unveil the mechanism that explains how and when the on-site interaction induces the effective spin-orbit interaction.

\begin{acknowledgments}
We thank S. Yunoki for helpful discussions. 
This work was supported in part by Grants for Excellent Graduate School, MEXT, Japan.
K. M. was supported by Grant-in-Aid for JSPS Fellows.
\end{acknowledgments}

\bibliography{submit.bib}
\end{document}